\documentclass[12pt]{article}
\usepackage{amsfonts,amsmath,amssymb,epsf}
\usepackage{graphicx,color}
\def\rishi{\rangle \! \rangle}

\def\prt{\partial}

\newcommand{\al}{\alpha'}

\newcommand{\be}{\begin{equation}}
\newcommand{\ba}{\begin{eqnarray}}
\newcommand{\ea}{\end{eqnarray}}
\newcommand{\ee}{\end{equation}}

\newcommand{\f}{\frac}
\newcommand{\s}{\sqrt}

\newcommand{\ti}{\tilde}
\newcommand{\ap}{\alpha}

\newcommand{\la}{\langle}
\newcommand{\lb}{\rangle}

\begin{document}

\begin{titlepage}
\thispagestyle{empty}
\begin{flushright}
hep-th/0309135\\
UT-03-28
\end{flushright}

\bigskip

\begin{center}
\noindent{\large \textbf{Boundary States for Supertubes in\\ \vspace{3mm} 
Flat Spacetime and G\"odel Universe
}}\\
\vspace{2cm}
\noindent{
Hiromitsu Takayanagi \footnote{hiro@hep-th.phys.s.u-tokyo.ac.jp}}
\\
\vskip 2.5em

{\it Department of Physics, Faculty of Science,
University of Tokyo\\
Hongo 7-3-1, Bunkyo-ku, Tokyo, 113-0033, Japan\\
\noindent{\smallskip}\\}

\vskip 2em
\end{center}

\begin{abstract}
We construct boundary states for supertubes
in the flat spacetime. The T-dual objects of supertubes
are moving spiral D1-branes (D-helices). Since we can obtain
these D-helices from
the usual D1-branes via null deformation, we can construct
the boundary states for these moving D-helices in the covariant formalism.
Using these boundary states, we calculate the vacuum amplitude
between two supertubes in the closed string channel and read
the open string spectrum via the open closed duality. We find
there are critical values of the energy for on-shell
open strings on the supertubes due to the non-trivial stringy correction.
We also consider supertubes in the type IIA G\"odel universe in order to use 
them as probes of closed timelike curves. 
This universe is the T-dual of the maximally supersymmetric type IIB
PP-wave background. Since the null deformations of D-branes are also allowed
in this PP-wave, we can construct the boundary
states for supertubes in the type IIA G\"odel universe
in the same way. We obtain the open
string spectrum on the supertube from the vacuum amplitude between
supertubes. As a consequence, we find that the tachyonic instability
of open strings on the supertube, which is the signal of closed time
like curves, disappears due to the stringy correction.

\end{abstract}
\setcounter{footnote}{0}
\end{titlepage}

\newpage

\section{Introduction}
\setcounter{equation}{0}
\hspace{5mm}
In the resent development of the string theory,
D-branes played an important role. Especially, since D-branes
are non-perturbative solitons, we have revealed
many non-perturbative aspects in the string theory
by studying the dynamics of D-branes.

Now although D-branes have tension, it is known that in a
type IIA string theory, tubular D2-branes in the flat
spacetime can be stable by adding proper electric and magnetic gauge
flux on them. These D2-D0-F1 bound states are called `supertubes'
\cite{MaTo}. Supertubes also preserve 8 supercharges in spite of its curved
configuration. An important feature of supertubes is
that we can interpret a supertube as a `blown up' object of D0-branes
(see \cite{BaLe} for the matrix description of D-particles).
Therefore it is interesting to study the blow up process from D0-branes
to supertubes. In this paper we would like to study this process
in the string theory.
Since the background RR-flux is not needed in contrast with the Myers'
effect \cite{My}, a dielectric effect of D-branes
in the presence of the background RR-flux,
there is a possibility to treat
this process in the NSR superstring formulation and indeed it is possible
as we will see.

Taking the T-duality in the direction in which the supertube is extended, 
we can obtain a moving spiral D1-brane (D-helix) from the supertube
as is claimed in \cite{ChOh}. The main idea used in this paper is that
we can get this moving D-helix from a static D1-brane
via `null deformation' \cite{CaKl,BaHu,MyWi,Ba}
(see also \cite{DuPi} for more generalized cases), the deformation
only in the $x^+$ direction. Therefore the blow up process
corresponds to the null deformation in the T-dual picture.

In this paper we would like to follow this null deformation
in the string theory using the boundary state formalism.
In the boundary state formalism, D-branes are treated as the
initial (final) states of closed strings and these states are
called boundary states. We will use this formalism because
the situation becomes simpler. While the time evolution of
open strings on supertubes is non-trivial
because of boundary interactions, closed strings propagate
trivially once they are emitted at the boundary.
Due to this advantage, we can treat the null deformation
of boundary states in the covariant formalism \cite{CaKl,HiTaTa}.
We express the null deformation (the blow up in the T-dual picture)
as deformation operators (Wilson loop) on the boundary
and as a consequence, we can obtain
boundary states for supertubes.
We would like to stress that we can treat these boundary states
exactly without light cone gauge fixing, i.e., without
breaking the conformal invariance of the string
world sheet. In the string theory, we highly depend on
the conformal invariance of the world sheet to calculate
string scattering amplitudes. Therefore we need to treat
strings covariantly for further analyses.

D-branes are also useful for studying stringy effects on geometries
because we can use them as probes \cite{DoMo}.
Since supertubes have tubular form, we can 
analyze stringy effects near a certain curve using supertubes as probes.
Especially they are useful in studying 
closed timelike curves (CTCs) in G\"odel universe \cite{Godel}.
The main motivation of studying G\"odel universe is to find out 
fates of backgrounds with CTCs and
it is interesting to study this problem in the string theory
\cite{GaBuHuPaRe,He,BoGaHoVa,HaTa,GiHa,DrFiSi,HiSJ,BrDePaRo,
HuRaRo,BrHeHi,Br,BrDaGrOl}.
In this sense, the supersymmetric type IIA G\"odel universe \cite{BoGaHoVa}
will be a good
example. This is because 
we can obtain the universe
via T-duality from the (compactified) maximally supersymmetric
type IIB pp-wave \cite{BlFiHuPa}
as is claimed in \cite{BoGaHoVa}.
Since we can construct an exactly solvable string theory on
the maximally supersymmetric
type IIB pp-wave background \cite{Me},
we can also treat the type IIA G\"odel universe in the string theory
\cite{HaTa}. This background also has 20 supersymmetries and we can expect
a high stability against quantum corrections.

In this paper we would also like to consider supertubes in
the supersymmetric type IIA G\"odel universe. We put
a supertube whose worldvolume is along a CTC
in order to use it as a probe
\cite{DrFiSi,HiSJ,BrHeHi,Br}. As before its T-dual
is a moving D-helix, a null deformed D1-brane,
in the the maximally supersymmetric type IIB PP-wave background.
As claimed in \cite{SkTa1}, the null deformations of D-branes are
also allowed in this PP-wave. We will
find that we can construct the boundary states for D-helices
in this background in the same way as in the flat spacetime.
We will compute the vacuum amplitude between supertubes
and obtain the open string spectra on them 
from which we can read the stringy correction to closed timelike
curves.

This paper is organized as follows. In section 2 we construct
the boundary states for the supertubes 
in the flat background and
consider the open string spectra on them.
We also construct the boundary
states for deformed supertubes \cite{MaNgTo1,MaNgTo2,HyOh}
(see also \cite{BaKa,BaOh,BaOhJa} for the supersymmetric
D2-anti D2 brane system which we can obtain by deforming a supertube). 
In section 3 we consider supertubes in type IIA G\"odel universe.
We construct boundary states for them in a similar manner and
obtain open string spectra on them. In section 4 we give a
brief summary of our results and draw conclusions.
\section{Boundary State for Supertube in Flat Spacetime}
\setcounter{equation}{0}
\hspace{5mm}
In this section we will denote the world volume directions
of the supertube as $(t,\phi,\tilde y)$. The supertube has
the gauge flux $F_{t\tilde y}=-1$ and $F_{\phi \tilde y}
=\mbox{constant}$. Though one might think the electric field is
at the critical value, it is not the case because of the non zero
magnetic flux.
To see the further property of supertubes,
let us compactify the $\tilde y$ direction with the radius $\tilde R$.
When a supertube consists of $N$ D2-branes, $k$ D0-branes and
$q$ F-strings, which we will call the $(N,k)$ supertube,
magnetic flux and the radius of the supertube denoted by $\rho$
are quantized as follows
\begin{equation}
F_{\phi \tilde y}=\frac{\al}{\tilde R}\frac{k}{N},\quad
\rho^2= g_s\f{|qk|\alpha^{\prime \f{3}{2}}}{N^2\tilde R},
\end{equation}
where $g_s$ is the coupling constant for closed strings.
The important feature is that the tension of this supertube equals
the sum of the tension of $|q|$ F-strings and $|k|$ D0-branes.
Therefore we can interpret supertube as the `blown up' D0-branes.

\begin{figure}[h!]
\begin{center}
\scalebox{0.7}{\includegraphics{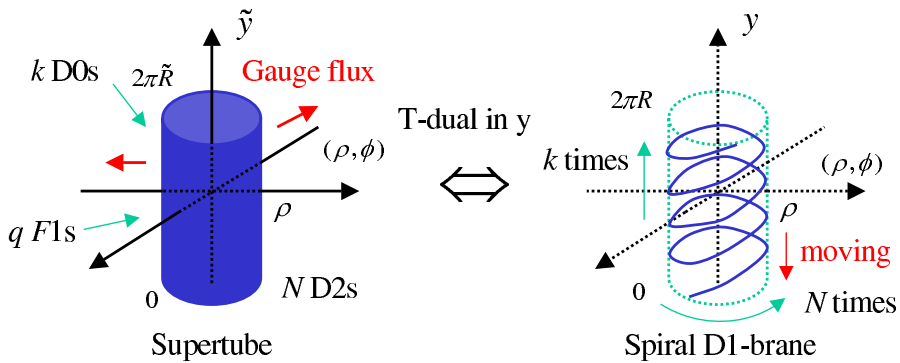}} \hspace{5mm}
\caption{Supertube and Spiral D1-brane
\label{Fig1}}
\end{center}
\end{figure}
Taking a T-duality in the $\tilde y$ direction, we can treat
the supertube more easily. As claimed in \cite{ChOh}, a T-dualized
object of the supertube is a moving `D-helix', a spiral D1-brane,
wound $k$ times in the $y$ direction and $N$ times
in the $\phi$ direction (see fig.\ref{Fig1}).
We can check this fact by treating D-branes as boundary conditions
of open strings. Although we will consider them in a bosonic
string theory for simplicity, we can easily extend obtained results
to super string cases as is discussed later.
Remembering that the (mixed) Neumann boundary condition is given by
\begin{equation}
G_{\mu\nu}\prt_N X^\nu +(B_{\mu\nu}+F_{\mu\nu})\prt_DX^\nu=0,
\label{mixed Neumann}
\end{equation}
we can write down the boundary conditions for the supertube as follows
\begin{equation}
\begin{split}
\mu=t:&\quad -\prt_N t -\prt_D \tilde y=0,\\
\mu=\phi:&\quad
+\rho^2 \prt_N \phi
+ b\prt_D \tilde y =0,\\
\mu=\tilde y:&\quad\prt_N \tilde y +\prt_D t -b\prt_D \phi =0,\\
\mu=\rho:&\ \quad \prt_D \rho=0,
\end{split}
\end{equation}
where we defined $F_{\phi \tilde y}=b$. Taking the T-duality
in $\tilde y$ direction and defining light cone directions as
$x^+=t+y$, $x^-=\frac{1}{2}(t-y)$, these change to
\footnote{In the limit of $b \rightarrow \infty$,
the condition (\ref{tube bc f}) becomes to that for
usual D1-branes extended in the $y$ direction. This is natural because
the supertube with the large magnetic flux is equivalent
to (infinite) D0-branes.}
\begin{equation}
\prt_N x^+=0, \quad \prt_N x^- -\frac{\rho^2}{b}\prt_N \phi=0,
\quad \prt_D (\phi -\frac{1}{b}x^+)=0
,\quad \prt_D \rho=0, \quad b=R\frac{k}{N},\label{tube bc f}
\end{equation}
where $R=\al/\tilde R$ is the radius of the y direction.
Eq.(\ref{tube bc f}) is just the conditions
for the moving D-helix in fig.\ref{Fig1}.

Now eq.(\ref{tube bc f}) tells us that we can obtain
this moving D-helix from usual D1-branes extended in
the $y$ direction via `null deformation'
\cite{CaKl,BaHu,MyWi,Ba,DuPi}, i.e.,
a deformation only in the $x^+$ direction. Noticing this fact
we would like to treat D-branes in the boundary state formalism
and construct the boundary states for supertubes via the null
deformation \cite{CaKl,HiTaTa}
in the next section.
First let us treat the case of $k=1$ for simplicity because the number
of segments of D1-brane is one \footnote{We will treat a D1-brane
wound $k$ times in the $y$ direction as $k$ segments of D1-branes.}.
\subsection{Boundary State for $k=1$ Supertube and Vacuum Amplitude}
As we have reviewed in the previous section, a supertube is described
as a moving spiral D1-brane (D-helix) in the T-dualized background.
Here we would like to
construct the boundary state for this D-helix
and calculate the vacuum amplitude.

Using the formalism in \cite{Ca}, we can get
the boundary state for the D-helix wound $(N,1)$
times in the $(\phi,y)$ directions (denoted as $|N,1\rangle$)
by multiplying one static D1-brane $|D1\rangle$ by
the proper Wilson loop-like operator. That is
\footnote{In this section we will use the mode expansion of
closed string in compact spacetime as
\ba
X^{\mu}(\tau,\sigma)
 =x^{\mu}+\al p^{\mu}\tau+R^\mu w_\mu \sigma
+i\s{\f{\al}{2}}\sum_{n\neq 0}\f{1}{n}
\Bigl(\ap^{\mu}_{n}e^{-in(\tau+\sigma)}
+\ti{\ap}^{\mu}_{n}e^{-in(\tau-\sigma)}\Bigr),\label{modee}
\ea
where the closed string $X^\mu(\tau,\sigma)$ 
has the periodicity under $\sigma \rightarrow \sigma +2\pi$. 
The commutation relations are
\begin{equation}
[\alpha^\mu_m,\alpha^\nu_n]=
[\tilde \alpha^\mu_m,\tilde \alpha^\nu_n]=
m\eta^{\mu\nu}\delta_{m,-n},\quad
[x^\mu,p^\nu]=i\eta^{\mu\nu},
\end{equation}
and the other commutators vanish. }
\begin{equation}
|N,1 \rangle = e^{-\frac{i}{2\pi \al}
\int_0^{2\pi}d \sigma V_{N,1}}|D1\rangle, \quad
V_{N,1}=\rho
\cos (\frac{NX^+}{R})\prt_\tau X^1
+\rho \sin(\frac{NX^+}{R})\prt_\tau X^2,
\end{equation}
where we used the relation $b=\frac{Rk}{N}$. Notice
that this boundary state preserves the boundary conformal invariance
including renormalization because the deformation is null \cite{CaKl}
(see also \cite{KrMyPeWi} for the proof for supertubes in the
context of the boundary conformal field theory).
We can also rewrite it using oscillators as follows
\begin{equation}
|N,1\rangle= \prod_{m=1}^\infty \exp
\Big[-\frac{m}{\al}V^i_mV^i_{-m}
-i\sqrt{\frac{2}{\al}}(V^i_m\ap^i_{-m}
+V^i_{-m}\tilde \ap^i_{-m})
\Big]|D1\rangle,
\end{equation}
where we defined
\begin{equation}
V^1_m(X^+) \equiv \frac{1}{2\pi}\int_0^{2\pi}d \sigma \rho
\cos(\frac{X^+}{b})
e^{im\sigma},\quad V^2_m(X^+)\equiv \frac{1}{2\pi}\int_0^{2\pi}d \sigma \rho
\sin(\frac{X^+}{b})
e^{im\sigma}. \label{vertex f}
\end{equation}
Notice that $\ap^+_{m}$'s commute each other. Then $V^i_m$ is
well defined.

Now let us compute the following vacuum amplitude between
the same $(N,1)$ D-helix
\begin{equation}
Z=\langle N,1|\Delta|N,1\rangle= \frac{\al}{2}\int_0^\infty d s
\langle N,1|e^{-sH}|N,1\rangle \equiv 
\int_0^\infty ds{\cal A}(s),
\end{equation}
where $\Delta$ is a closed string propagator and the function ${\cal A}(s)$
is given by
\begin{equation}
{\cal A}(s)=\frac{\al}{2}\langle N,1|e^{-sH}|N,1\rangle=
\frac{\al}{2}\langle D1|e^{\frac{i}{2\pi \al}
\int_0^{2\pi}d \sigma V_{N,1}}e^{-sH}e^{-\frac{i}{2\pi \al}
\int_0^{2\pi}d \sigma V_{N,1}}|D1\rangle.
\end{equation}
As is shown in \cite{HiTaTa}, we can neglect massive modes in the
$x^+$ direction, i.e., $\ap^+_m$ for $m\neq 0$, in evaluating
this amplitude. Considering only zero modes,
$V^i_m$ has the following simple form
\begin{equation}
V^{1}_{\pm Nw}=\frac{\rho}{2}e^{\mp i\frac{Nx^+}{R}},\quad
V^{2}_{\pm Nw}=\pm i\frac{\rho}{2}e^{\mp i\frac{Nx^+}{R}},\quad
\mbox{otherwise zero},
\end{equation}
where $w$ denotes the winding number in the $y$ direction,
i.e., $X^+=x^++Rw\sigma +\cdots$.
Notice that this simpleness is due to the periodicity of the function
$V^{i}_{N,1}$ under the shift $X^+ \rightarrow X^+ +2\pi R$.
Then we can use the following effective boundary state in calculating
the vacuum amplitude
\begin{equation}
|N,1\rangle \sim \exp
\Big[-\frac{|Nw|}{\al}V^i_{|Nw|}V^i_{-|Nw|}
-i\sqrt{\frac{2}{\al}}(V^i_{|Nw|}\ap^i_{-|Nw|}
+V^i_{-|Nw|}\tilde \ap^i_{-|Nw|})
\Big]|D1\rangle.
\end{equation}
Employing the formula
\begin{equation}
\begin{split}
&\la 0|e^{-\f{1}{m}\ap_{m}\ti{\ap}_{m}}
e^{(f^{(2)}_{-m}\ap_{m}+f^{(2)}_m\ti{\ap}_{m})}
e^{-s(\ap_{-m}\ap_m + \tilde \ap_{-m}\tilde \ap_m)}
e^{(f^{(1)}_{m}\ap_{-m}+f^{(1)}_{-m}\ti{\ap}_{-m})}
e^{-\f{1}{m}\ap_{-m}\ti{\ap}_{-m}}|0\lb\\
&\ \ =\f{1}{1-e^{-2ms}}
\exp\Bigl(-\f{m}{e^{2ms}-1}\bigl[|f^{(1)}_{m}|^2+|f^{(2)}_{m}|^2
-e^{ms}(f^{(1)}_{m}f^{(2)}_{-m}+f^{(1)}_{-m}f^{(2)}_{m})\bigr]\Bigr)
\label{BHCg},
\end{split}
\end{equation} 
which can be shown by repeatedly applying the Baker-Campbell-Hausdorff's 
formula ($f^{(1,2)}_{m}(=f^{(1,2)*}_{-m})$ 
are any constants), we obtain the following vacuum amplitude
\begin{equation}
{\cal A}(s)=\frac{{\cal N}}{(2\pi\al s)^{12}}
\sum_w \exp \Big[-\frac{\rho^2}{\al}Nw \tanh
\frac{Nws}{2}\Big]
e^{-\frac{R^2w^2s}{2\al}}\eta(is/\pi)^{-24},
\label{vac supertube f} 
\end{equation}
where $\eta(is/\pi)^{-24} \equiv
\f{e^{2s}}{ \prod_{m=1}^{\infty}(1-e^{-2ms})^{24}}$ is the usual
$\eta$ function and the normalization factor ${\cal N}$ is given by
${\cal N}=\frac{\al T_1^2}{8}V_1$ \footnote{We defined the normalization
constant $T_p$ for Dp-brane as
$T_p\equiv 2^{7-p}\pi^{\frac{23}{2}-p}\alpha^{\prime \frac{11-p}{2}}$
and $V_1$ is the volume of the light cone directions.}. After taking
the T-duality in the
$y$ direction,
i.e., $R \rightarrow \al/\tilde R$ noticing $N=R/b$,
we get the desired vacuum amplitude between the same supertube.

Finally we would like to extend this result to superstring cases.
As is discussed in \cite{HiTaTa}, 
massive modes are decoupled similarly, and
we only have to replace the modular function 
 $\eta(is/\pi)^{-24}$ by the familiar terms with theta-functions
\ba
\f{\theta_3(is/\pi)^4-\theta_2(is/\pi)^4-\theta_4(is/\pi)^4}
{2\eta(is/\pi)^{12}}(=0).
\label{theta}
\ea
Therefore the vacuum amplitude vanishes. This is 
consistent with the fact that a supertube preserves
eight supersymmetries. We can also get a non trivial
result considering a `tube-antitube' system, i.e., a tubular D2-anti D2
system with the same gauge flux on both branes. The vacuum amplitude
can be obtained by just changing the sign in front of $\theta_2(is/\pi)^4$
in (\ref{theta}). In this case the amplitude does not vanish
and we can see a non-trivial effect.
\subsection{Open Closed Duality in Discretized LC Gauge}
In the previous section we have evaluated the vacuum amplitude
between the same D-helix. Before reading an open string spectrum
from this amplitude, we would like to review the open-closed
duality in the discretized LC gauge \footnote{See \cite{BeGaGr,GaGr}
for open closed duality of D-instantons in the light cone gauge.}
in this section. We will obtain
an exact open string spectrum on the D-helix in the LC gauge using
a technique shown in this section.

Let us consider the open string 1 loop amplitude in the light cone
gauge with the Lorentzian world-sheet.
Since the $y$ direction
is compactified with the radius $R$ in our case, the momentum is discretized
as
\begin{equation}
p^+ =E+\frac{n}{R},\quad p^-=\frac{1}{2}(E-\frac{n}{R}).
\end{equation}
Denoting the open string (Euclidean) modular parameter $t$ as
$\pi/s$, the open string 1 loop amplitude in this gauge
is given by 
\begin{equation}
\begin{split}
Z_{open}
=&2\times V_1\int_0^\infty \frac{dt}{2t}
\int \frac{d p^+}{(2\pi)^2R} \sum_n \mbox{Tr}\, \exp
\Big[-2\pi t\Big(-\al p^+(p^+-\frac{2n}{R})+H_{o}\Big)\Big]\\
=&\frac{V_1}{8\pi^2 \al}\int_0^\infty \frac{dt}{t^2}\sum_w
\mbox{Tr}\, e^{-2\pi t (-\al p^{+2}+H_{o})}|_{p^+=\frac{wR}{2i\al t}},
\label{open 1loop}
\end{split}
\end{equation}
where in the second line we performed integral
\footnote{We used the following resummation formula
\begin{equation}
\sum_{n=-\infty}^\infty e^{inx}= 2\pi \sum_{w=-\infty}^\infty
\delta(x-2w\pi).
\end{equation}}
with respect to the light cone directions
in a similar manner as in \cite{DuPi}.
Notice that the modular parameter $t$ is Euclidean
while the world-sheet is Lorentzian. Therefore 
we should treat $t$  as a pure imaginary one
in performing the integral in eq.(\ref{open 1loop}).
We would like to interpret this result as
the closed string tree amplitude. Considering
the Euclidean world-sheet, i.e., treating $t$ as a real one
in the second line
in eq.(\ref{open 1loop}), the second line
corresponds to taking the following gauge in the
open string channel
\begin{equation}
X^+ =2\al p^+\tau=\frac{wR}{it}\tau=\frac{wR}{t}\tau_E,
\end{equation}
where $\tau_E$ is the Euclidean time. Under the open closed
duality $\tau_E \leftrightarrow t\sigma$, 
this changes to the gauge $X^+=Rw\sigma$
in the closed string channel. Therefore we can interpret
$w$ in eq.(\ref{open 1loop}) as a winding number of a closed string
in the $y$ direction.

Now the main claim in this section is that we can obtain the result
in the closed string channel by replacing $2i\al p^+t$ in the
open string channel with $wR$ 
and vise versa. For example, the first factor in eq.(\ref{open 1loop})
corresponds to the effect of winding modes as in eq.(\ref{vac supertube f});
\begin{equation}
e^{2\pi t \al p^{+2}}|_{p^+=\frac{wR}{2i\al t}}
=e^{-\frac{w^2R^2\pi}{2\al t}}=e^{-\frac{w^2R^2s}{2\al}}
\label{example}.
\end{equation}
Finally noticing that there is no momenta in the light cone directions
in the closed string channel, we find the Cardy's condition
is given by
\begin{equation}
\mbox{Tr}\, e^{-2\pi t H_{o}}|_{p^+=\frac{wR}{2i\al t}}
=\langle B|e^{-sH_{c}}|B\rangle|_{X^+=wR\sigma}. \label{cardy}
\end{equation}
\subsection{Open string spectrum on the $k=1$ Supertube}
Let us now return to the problem of evaluating the open string spectrum.
Replacing $wR$ with $2i\al p^+ t$, we can easily rewrite
the non-trivial exponential factor in eq.(\ref{vac supertube f})
in the language of the open string channel as eq.(\ref{example}). That is
\begin{equation}
\exp \Big[-\frac{\rho^2}{\al}Nw \tanh
\frac{Nws}{2}\Big] \quad \rightarrow \quad
\exp\Big[-2\pi t\Big(-\frac{\rho^2}{b}\frac{p^+}{\pi}
\tan \frac{\al \pi p^+}{b}\Big)\Big],\label{ex nontri}
\end{equation}
where we used a relation $N=R/b$. Then we find the open string
Hamiltonian on the $(N,1)$ D-helix has the following form
\begin{equation}
H_o=-2\al p^+p^- -\frac{\rho^2}{b}\frac{p^+}{\pi}
\tan \frac{\al \pi p^+}{b}+H_{osc},
\end{equation}
where $H_{osc}$ denotes a contribution of massive modes, i.e.,
\begin{equation}
H_{osc}=\sum_{i=1}^{24}
\sum_{m=1}^\infty \ap^{i}_{-m}\ap^i_m-1.
\end{equation}
After taking T-duality in the $y$ direction, we get the desired
open string Hamiltonian on the $(N,1)$ supertube as follows
\begin{equation}
H_o=-\al E^2 +\frac{\tilde R^2 w^2}{\al}
-\frac{\rho^2}{b}\frac{1}{\pi}(E+\frac{\tilde R w}{\al})
\tan\Big[\frac{\al \pi}{b}(E+\frac{\tilde R w}{\al})\Big]+H_{osc}
\label{open supertube f}.
\end{equation}
We can also obtain the result in a superstring theory,
by replacing $H_{osc}$ with that for a superstring theory.

Now we would like to carry out physical interpretations from this result.
We will treat only
no winding open strings ($w=0$) for simplicity.
If we consider the low energy limit $|\al E/b| \ll 1$,
we can expand the open string Hamiltonian as follows
\begin{equation}
H_o=-\al E^2 (1+\frac{\rho^2}{b^2}) + \cdots.
\end{equation}
We can interpret the result as the usual expression
$H_o=\al p_\mu p_\nu G_{eff}^{\mu \nu}+\cdots$ with
the open string metric $G_{eff}^{\mu \nu}$ \footnote{
the open string metric is defined as (see \cite{SeWi} in detail)
\begin{equation}
G_{eff}^{\mu\nu}=\left(\frac{1}{G+B+F}\right)_S^{\mu\nu},
\end{equation}
where the symbol $S$ denotes the symmetric part of a matrix.}
and is indeed consistent with the open string metric for a supertube
$G^{tt}_{eff}=-(1+\f{\rho^2}{b^2})$.

On the other hand, in the high energy region we can see a non-trivial
$\al$ correction in the Hamiltonian (\ref{open supertube f}). The
open string spectrum ($H_o=0$) is given by
\begin{equation}
\al E^2
+\frac{\rho^2}{b}\frac{E}{\pi}
\tan\Big[\frac{\al \pi E}{b}\Big]=M^2.
\end{equation}
Notice that there is always a solution for any $M^2\geq 0$
in the range of $0 \leq E<\frac{|b|}{2\al}$. Therefore
we find that there is a `critical' energy for
open strings. 

Finally let us comment on several points before finishing this section.
First we cannot take the large radius limit $\tilde R\rightarrow \infty$
because in this limit $b$ goes to zero and the obtained results
become subtle. This is physically natural because
a $(N,1)$ supertube is the blown up object of one D0 brane.
In order to keep the D0 charge density finite in the large
radius limit, we need to consider cases of $k \neq 1$
(especially $k=\infty$).

We can also check the
Cardy's condition for D-branes with any traveling waves \cite{HiTaTa},
\begin{equation}
|P\rangle =\exp\Big[-\frac{i}{2\pi \al}\int_0^{2\pi}d \sigma
\Phi_i(X^+)\prt_\tau X^i\Big]|D1\rangle,
\end{equation}
using the same technique as in eq.(\ref{example}).
Due to the periodicity of waves with the shift
$X^+ \rightarrow X^+ +2\pi R$, the non trivial factor as
in eq.(\ref{ex nontri}) is always expressed as $\exp[wf(ws)]$
with a proper function $f$. Therefore we can obtain
the result in the open string channel with the
replacement to $\f{2i\al p^+t}{R}$ from $w$ as follows
\footnote{For example we can rewrite
the exponential factor between D-branes with the same traveling waves
(see the section 3 in \cite{HiTaTa} in detail) as follows
\begin{equation}
\exp\Big[-\sum_{n=1}^\infty n|c^i_n|^2 \f{2w}{\al}\tanh \f{nws}{2}\Big]
\quad \rightarrow \quad \exp\Big[-2\pi t\Big(-\sum_{n=1}^\infty
n|c^i_n|^2 \f{2p^+}{\pi R}\tan (\f{\pi n \al p^+}{R})\Big)\Big],
\end{equation}
where $|c^i_n|$ is the Fourier mode of $\Phi^i(X^+)$ defined as in
eq.(\ref{vertex f}).}
\begin{equation}
\exp[wf(ws)] \rightarrow \exp\Big[\f{2i\al p^+ t}{R}
f(\f{2i\pi \al p^+}{R})\Big]=\exp[-2\pi t(p^+\tilde f(p^+))].
\end{equation}
There is no extra normalization factor and
the Cardy's condition is satisfied as expected.
\subsection{Boundary State for $k\neq 1$ Supertube}
\begin{figure}[h!]
\begin{center}
\scalebox{0.7}{\includegraphics{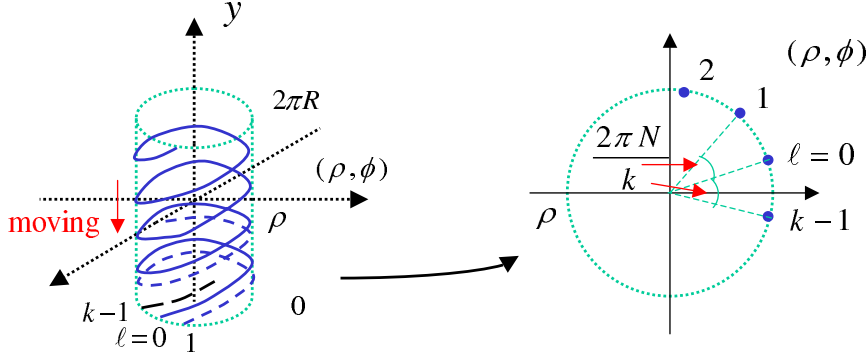}} \hspace{5mm}
\caption{$(N,k)$ D-helix from k segments of spiral D1-branes
\label{Fig2}}
\end{center}
\end{figure}

As found in the previous section, we can not take the
large radius limit $\tilde R \rightarrow \infty$
for $k=1$ supertubes. In this section 
let us consider the $k \neq 1$ supertubes in order to treat
this limit. As discussed before, we can interpret one $(N,k)$
D-helix, the T-dual object of a $(N,k)$ supertube, as symmetrically
placed $k$ segments of spiral D1-branes (see fig.\ref{Fig2}).

Then we can construct the boundary state for the $(N,k)$ D-helix
(denoted as $|N,k\rangle$) by superposing k boundary states for
properly shifted spiral D1-branes. That is
\begin{equation}
|N,k \rangle = \sum_{\ell=0}^{k-1} |N,k;\ell\rangle,\quad
|N,k;\ell\rangle \equiv
e^{-\frac{i}{2\pi \al}\int_0^{2\pi}d \sigma V_{N,k;\ell}}|D1\rangle,
\end{equation}
where we defined the deformation operator $V_{N,k;\ell}$ as follows
\begin{equation}
V_{N,k;\ell} \equiv \rho
\cos \Big(\frac{NX^+}{kR}+2\pi \frac{N\ell}{k}\Big)\prt_\tau X^1
+\rho\sin\Big(\frac{NX^+}{kR}+2\pi \frac{N\ell}{k}\Big)\prt_\tau X^2.
\label{deform op st}
\end{equation}
We can interpret one segment as a `fractional' brane
and the $(N,k)$ D-helix as a `bulk' brane in an orbifold theory
\footnote{We can find the similar property in D-branes
in Melvin background \cite{DuMo,TaUe} (see also \cite{TaTa}
for D-branes in the exactly solvable model \cite{RuTs} (a time dependent
version of Melvin background)).}. This is because we can obtain
a $(N,k)$ D-helix from a $(N,1)$ D-helix in the space whose
the radius in the $y$ direction is $kR$ via ${\Bbb Z}_k$ orbifolding.
In the context of the orbifold theory, we can understand
only one segment or any D-branes which are not periodic
with $X^+\rightarrow X^+ +2\pi R$ are inconsistent
because these are not invariant under the ${\Bbb Z}_k$ symmetry.

Now let us compute the vacuum amplitude. Although we can
easily calculate it in the same way as appeared in the
previous section, the obtained result is not simple this time
since the deformation operator $V_{N,k;\ell}$ is not
periodic with $X^+ \rightarrow X^+ +2\pi R$ any longer.
Anyway we find the following amplitude for $(N,k)$ D-helix
\begin{equation}
{\cal A}(s)=\f{{\cal N}}{(2\pi \al s)^{12}}
\sum_{\ell,\ell'=0}^{k-1}\sum_{w=-\infty}^\infty
\exp\Big[-\sum_{m=1}^\infty L^{(\ell,\ell')}_{m,w}(s)\Big]
e^{-\frac{R^2w^2s}{2\al}}\eta^{-24}(is/\pi),
\end{equation}
where the non trivial factor for the amplitude between $|N,k;\ell\rangle$
and $|N,k;\ell'\rangle$ is given by
\begin{equation}
L^{(\ell,\ell')}_{m,w}(s)=\frac{2\rho^2}{\pi^2\al}
\frac{(m^2+(\frac{Nw}{k})^2)}{(m^2-(\frac{Nw}{k})^2)^2}\sin^2\Big(
\frac{\pi Nw}{k}
\Big)\frac{m}{e^{2ms}-1}
\Big[e^{2ms}+1-2e^{ms}\cos\Big(\frac{2\pi N}{k}(\ell-\ell')\Big)\Big].
\end{equation}
This amplitude reduces to eq.(\ref{vac supertube f}) in the limit
of $k\rightarrow 1$ as expected. After taking a T-duality in
the $y$ direction, we obtain the desired amplitude for the $(N,k)$
supertube.

Especially, we are interested in the large radius limit
$\tilde R\rightarrow \infty$ of this amplitude. To keep
the D0 charge density finite, we need to take the following limit
\begin{equation}
k \rightarrow \infty, \quad \f{\tilde R}{\alpha^{\prime 1/2}}\rightarrow
\infty, \quad \mbox{with}\quad b=\frac{\al}{\tilde R}\f{k}{N}:\mbox{fixed}.
\end{equation}
Notice that this corresponds to the deconstruction limit
of the ${\Bbb Z}_k$ orbifold \cite{ArCoGe}
(see the right picture of fig.\ref{Fig2}). In this limit, we can
express the vacuum amplitude for the $(N,k)$ supertube as follows
\begin{equation}
{\cal A}(s)=
k^2{\cal N}_0\int_0^1 d x \int_{-\infty}^\infty d p 
\f{1}{(2\pi \al s)^{12}}
\exp\Big[-\sum_{m=1}^\infty L_m(p,x,s)\Big]
e^{-\frac{R^2w^2s}{2\al}}\eta^{-24}(is/\pi),
\end{equation}
where we defined ${\cal N}_0$ as a usual normalization factor
for D0-brane, i.e., ${\cal N}_0=\f{\al T_0^2}{8}V_0$ and
the non trivial exponential factor is given by
\begin{equation}
L_m(p,x,s)=\f{2\rho^2}{\pi^2 \al}\f{m^2+(\f{\al p}{b})^2}
{\big(m^2-(\f{\al p}{b})^2\big)^2}\sin^2 (\f{\pi \al p}{b})\f{m}{e^{2ms}-1}
(e^{2ms}+1-2e^{ms}\cos(2\pi Nx)).
\end{equation}
Although it seems to be difficult to rewrite this amplitude
in the language of the open string channel because we can not use
the technique as in eq.(\ref{example}), it will be interesting
to compare above results with the matrix description of
D-particles \cite{BaLe}.

\subsection{Generalization}
\begin{figure}[h!]
\begin{center}
\scalebox{0.7}{\includegraphics{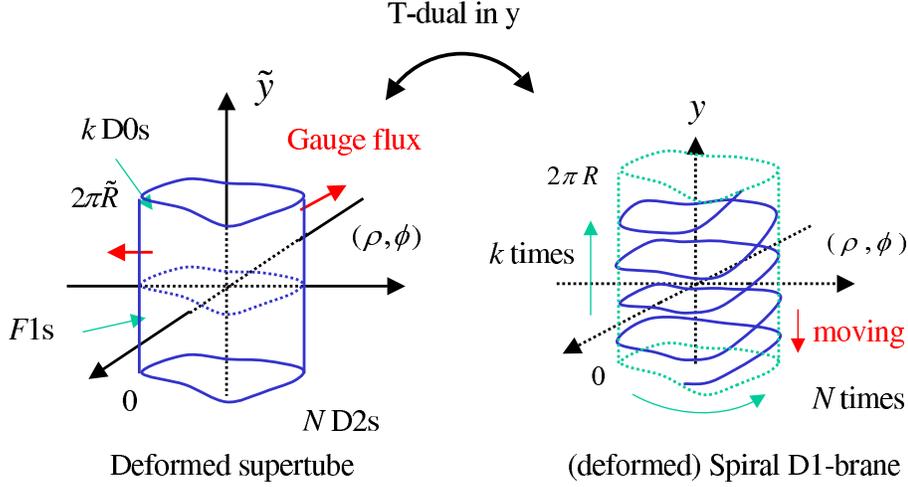}} \hspace{5mm}
\caption{Deformed supertube and deformed D-helix
\label{Fig3}}
\end{center}
\end{figure}

As is claimed in \cite{MaNgTo1}, we can deform
supertubes preserving 8 supersymmetries. Although
the shapes of the sections of a tube in the constant
$\tilde y$ plane have
to be the same for any $\tilde y$, any shapes of section
are OK (see the left picture in fig.\ref{Fig3})\footnote{In \cite{MaNgTo2},
it is claimed that a deformed supertube is equivalent to a `supercurve',
a (deformed) moving spiral F-string, under the TST duality transformation.
We can also understand this fact in the context of the M-theory
as the 9-11 flip of a `M-ribbon' \cite{HyOh}, a spiral M2-brane.}. 
Making T-duality in the $\tilde y$ direction, we
obtain from a deformed $(N,k)$ supertube a deformed moving $(N,k)$ D-helix
wrapped on the surface of the deformed supertube as claimed in \cite{HyOh}
(see fig.\ref{Fig3}). In this T-dual picture, we can easily understand
the validity of this deformation from the point of view of null deformation.

Therefore when the section with the plane $\tilde y = \mbox{const.}$ has
the following form
\begin{equation}
X^1=\rho(\phi)\cos \phi, X^2=\rho(\phi)\sin \phi,
\end{equation}
we can construct the boundary state for the deformed $(N,k)$ D-helix,
the T-dual object of the deformed $(N,k)$ supertube,
by using the following deformation operator instead of eq.(\ref{deform op st})
\begin{equation}
\begin{split}
V_{N,k;\ell} \equiv \,\,&\rho\Big(\frac{NX^+}{kR}+2\pi \frac{N\ell}{k}\Big)
\cos \Big(\frac{NX^+}{kR}+2\pi \frac{N\ell}{k}\Big)\prt_\tau X^1\\
&+\rho\Big(\frac{NX^+}{kR}+2\pi \frac{N\ell}{k}\Big)
\sin\Big(\frac{NX^+}{kR}+2\pi \frac{N\ell}{k}\Big)\prt_\tau X^2.
\end{split}\label{deform op sc}
\end{equation}
In the same way as before, we can obtain exact vacuum amplitudes
for deformed supertubes in the closed string channel, which vanish
in the superstring case as expected from the point of view of
the supersymmetry.
Especially we can get the exact open string spectrum for
deformed $(N,1)$ supertubes using the same technique
as in eq.(\ref{example}).
\section{Supertube in Type IIA G\"odel Universe}
\setcounter{equation}{0}
\hspace{5mm}
In this section we would like to consider supertubes
on the type IIA G\"odel universe \cite{BoGaHoVa}, which is the T-dual
of the type IIB PP-wave background with the maximal supersymmetry
\cite{BlFiHuPa}, in order to use them as probes of closed timelike curves.
The supertubes correspond to moving D-helices, null deformed D1-branes,
in the PP-wave via T-duality as in the case of the flat spacetime.
As claimed in \cite{SkTa1}, the null deformed D-branes satisfy
equation of motions also in the case of this PP-wave.
We will find that we can construct
the boundary states for supertubes in the universe in the same way as
in the flat spacetime.

\subsection{Relation between the G\"odel Universe and PP-Wave}
In this section we would like to give a brief review of
the relation between the type IIA G\"odel universe
and the maximally supersymmetric type IIB PP-wave background.
The type IIA G\"odel universe has the following metric
\begin{equation}
ds^2=-(dt+\beta \sum_{i,j=1}^8
J_{ij}x^idx^j)^2+\sum_{i=1}^8(d x^i)^2+d\tilde y^2,
\quad J_{2k-1,2k}=s_k=\pm 1,
\label{godel}
\end{equation}
and NS-NS and RR 4-form flux
\begin{equation}
H_{2k-1,2k,\tilde y}=-2s_k\beta ,\quad
F_{1234}=F_{5678}=4\beta .
\label{godel flux}
\end{equation}
We will use the complex coordinates $z^k$ and
the polar coordinates $(\rho_k,\phi_k)$ defined as
\begin{equation}
z^k= \rho_k e^{i\phi_k}\equiv x^{2k-1}+is_k x^{2k}.\label{cpx}
\end{equation}
This background has the closed time like curves for
$\rho_k > 1/\beta$. 

On the other hand, the type IIB maximally supersymmetric PP-wave
background is given by
\begin{equation}
\begin{split}
ds^2 = -2 dx^+ dx^- -\beta^2 \sum_{i=1}^8(\tilde x^i)^2(dx^+)^2 +
\sum_{i=1}^8(d\tilde x^i)^2,\\
F_{(5)}=4\beta dx^+ (d \tilde x^1
d\tilde x^2 d\tilde x^3 d\tilde x^4
+d \tilde x^5
d\tilde x^6 d\tilde x^7 d\tilde x^8),
\end{split}\label{ppw}
\end{equation}
where we defined the light cone directions as $x^+=t+y$, $x^-=(t-y)/2$
as before. The coordinate transformation
\begin{equation}
\tilde x^{2k-1}+ is_k \tilde x^{2k}=(x^{2k-1}+is_k x^{2k})e^{-i\beta x^+}
\end{equation}
changes this background into the following one
\begin{equation}
ds^2=-dt^2+dy^2+\sum_{i}(d x^i)^2-2\beta
\sum_{ij}J_{ij}x^idx^j(dt+dy).
\label{twist ppw}
\end{equation}
In this article we will call this background `the twisted PP-wave background'.
Taking T-duality in $y$ direction, we obtain the G\"odel universe
(\ref{godel}),(\ref{godel flux}).

The twisted PP-wave background has
the following symmetry (the translation in $x^i$)
that leaves the metric and the flux unchanged
\begin{equation}
z^k \rightarrow z^k-\rho_{0k}e^{i\phi_{0k}},\quad
x^- \rightarrow x^- + \sum_{k=1}^4
\beta \rho_{0k} \rho_k\sin(\phi_k-\phi_{0k}).
\label{trans}
\end{equation}
This symmetry is equivalent to that for the PP-wave background 
(\ref{ppw}) discussed in \cite{SkTa1}. In this paper we would like to
consider only the case of $s_k=(1,1,1,-1)$ because
the G\"odel universe has 20 supersymmetries in this case.

\subsection{Supertube on the G\"odel universe}
Now let us consider
the supertube extended in $(t,\phi_1,\tilde y)$
in the type IIA G\"odel universe. As in the case of
the flat spacetime, the analysis of the DBI action
\footnote{In spite of the presence of the RR 4-form flux in this universe,
we can find there is no Chern-Simons term for this supertube.} shows
a supertube with the gauge flux $F_{t\tilde y}=-1$
and $F_{\phi_1 \tilde y}=\mbox{constant}$ is a stable object.
In order to take the T-duality in the $\tilde y$ direction, let us
compactify the $\tilde y$ direction with the radius $\tilde R$.
As in the case of the flat spacetime, the magnetic flux and
the radius of the supertube $\rho$ are quantized as follows
\begin{equation}
F_{\phi_1 \tilde y}=
\frac{\al}{\tilde R}\frac{k}{N},\quad
\rho^2= \f{g_s\alpha^{\prime \f{1}{2}}}{N}
\left|\f{qF_{\phi_1 \tilde y}}{1-F_{\phi_1 \tilde y}\beta}\right|,
\end{equation}
where $N,k$ and $q$ are the numbers of D2-branes, D0-branes and F-strings
respectively. In this paper we would like to treat
only the case of $F_{\phi_1 \tilde y}\beta <1$ because
the tension of the supertube is the same as the sum of the tension
of $|k|$ D0-branes and $|q|$ F-strings only in the case
\footnote{For $F_{\phi_1 \tilde y}>1/\beta$, the tension of
the supertube is the tension of $|k|$ D0-branes minus the
tension of $|q|$ F-strings and
$\rho$ goes to infinity for $F_{\phi_1\tilde y}=1/\beta$.}.

We can also show that the T-dual of this supertube is
a moving spiral D1-brane as before.
We can express the supertube as the following boundary conditions of
open strings
\begin{equation}
\begin{split}
\mu=t:&\quad -\prt_N t -\beta \rho^2 \prt_N \phi -\prt_D \tilde y=0,\\
\mu=\phi:&\quad
-\beta \rho^2 \prt_N t +(\rho^2 -\beta^2\rho^4)\prt_N \phi
+ (b-\beta \rho^2)\prt_D \tilde y =0,\\
\mu=\tilde y:&\quad\prt_N \tilde y +\prt_D t +(\beta\rho^2-b)\prt_D \phi =0,\\
\mu=\rho:&\ \quad \prt_D \rho=0,
\end{split}
\end{equation}
where we defined $F_{\phi \tilde y}=b$. Taking the T-duality
in $\tilde y$ direction, these change to 
\begin{equation}
\prt_N x^+=0, \quad \prt_N x^-+(\beta-\frac{1}{b})\rho^2\prt_N \phi=0,
\quad \prt_D (\phi -\frac{1}{b}x^+)=0
,\quad \prt_D \rho=0, \label{tube bc}
\end{equation}
where we used the following relations of the T-duality
(see \cite{HaTa} in detail)
\begin{equation}
\prt_N \tilde y=\prt_D y -\beta \rho^2 \prt_D  \phi,\quad
\prt_D \tilde y=\prt_N y -\beta \rho^2 \prt_N \phi.
\end{equation}
After the translation (\ref{trans}),
we find that eq.(\ref{tube bc}) is just the condition of
the moving spiral D1-brane in the twisted PP-wave background.

\subsection{String Theory on the Twisted PP-wave Background}
In the previous section we have found that as in the case of the flat
spacetime, the T-dual of the supertube in the type IIA G\"odel universe
is a moving spiral D1-brane in the twisted PP-wave background.
Before constructing the boundary state for this spiral D1-brane,
let us review the string theory
on the twisted PP-wave background (\ref{twist ppw}) (see \cite{HaTa}
in detail). As expected from the name of the background,
this is just the twisted version of the string theory
on the type IIB maximally supersymmetric PP-wave background \cite{Me}.
In this section we will treat only the bosonic part
because the (GS) fermionic part is irrelevant to our analysis.
We will review closed strings first, consider
open strings on the static D1-brane extended in $t$ and $y$ directions
next and finally study the open closed duality of this D1-brane.

\subsubsection*{Closed String Theory}
Now let us consider the closed string theory on the twisted PP-wave.
Since the $y$ direction is compactified with the radius $R=\al/\tilde R$,
we should impose the light cone gauge $X^+ = \al p^+ \tau +wR\sigma$.
In this gauge, the (light cone) closed string action is given by
\begin{equation}
S_{closed}=-\frac{1}{2\al}\int d \tau \int_0^{2\pi} d\sigma
[\prt_a X^i \prt^a X^i
+2\beta X^i J_{ij}(p^+\prt_\tau X^j-wR\prt_\sigma X^j) ],\label{bulk action}
\end{equation}
where we set the world sheet metric at $\eta^{ab}=\mbox{diag}(-1,1)$.
After rewriting this action with the complex coordinates $Z^k$
in eq.(\ref{cpx}) and defining $Z^k_0$
as $Z^k=e^{i\beta s_k X^+}Z^k_0$, we find that 
the equation of motion of $Z^k_0$ is just that of the type IIB maximally
supersymmetric PP-wave background. That is
\begin{equation}
(\prt_\tau^2 -\prt_\sigma^2) Z^k_0+f^2Z^k_0=0,\quad f^2
=(\beta\al p^+)^2-(\beta Rw)^2.
\end{equation}
Noticing that closed strings should satisfy the following
twisted boundary condition
\begin{equation}
Z^k_0(\tau,\sigma+2\pi)=e^{-2\pi i\delta}Z^k_0(\tau,\sigma),
\quad \delta =s_k\beta Rw,
\end{equation}
we find the mode expansions of $Z^k_0$ and $\bar Z^k_0$, the complex
conjugate of $Z^k_0$, are given by
\begin{equation}
\begin{split}
Z^k_0&= i\sqrt{\frac{\al}{2}}\sum_{n \in {\Bbb Z}}
\Big(\frac{\ap^{k}_{n+\delta}}{w^+_n}
e^{-iw^+_n\tau-i(n+\delta)\sigma}
+\frac{\tilde \ap^{k}_{n-\delta}}{w^-_n}
e^{-iw^-_n\tau+i(n-\delta)\sigma}\Big),\\
\bar Z^k_0&= i\sqrt{\frac{\al}{2}}\sum_{n \in {\Bbb Z}}
\Big(\frac{\bar \ap^{k}_{n-\delta}}{w^-_n}
e^{-iw^-_n\tau-i(n-\delta)\sigma}
+\frac{\bar{\tilde \ap}^{k}_{n+\delta}}{w^+_n}
e^{-iw^+_n\tau+i(n+\delta)\sigma}\Big),
\end{split}\label{mode exp}
\end{equation}
where we defined 
\begin{equation}
\begin{split}
&w^+_n = \left\{\begin{array}{ll}
\sqrt{(n+\delta)^2+f^2} & \mbox{for} \quad n \geq -\delta \\
-\sqrt{(n+\delta)^2+f^2} & \mbox{for} \quad n < -\delta
\end{array}\right.,\\
&w^-_n = \left\{\begin{array}{ll}
\sqrt{(n-\delta)^2+f^2} & \mbox{for} \quad n > \delta \\
-\sqrt{(n-\delta)^2+f^2} & \mbox{for} \quad n \leq \delta
\end{array}\right..
\end{split}
\end{equation}
Oscillators satisfy the following quantization condition
\begin{equation}
[\ap^k_{n+\delta},\bar \ap^{k'}_{\ell-\delta}]=2w^+_n\delta_{\ell+n,0}
\delta_{kk'},\quad
[\tilde \ap^k_{n-\delta},\bar {\tilde \ap}^{k'}_{\ell+\delta}]
=2w^-_n\delta_{n+\ell,0}\delta_{kk'}.
\end{equation}
The closed string Hamiltonian is expressed as follows
\begin{equation}
{\cal H}_c=\frac{R^2w^2}{2\al}
-\al p^+p^-+H_{c}.
\end{equation}
The massive mode $H_c$ is given by
\begin{equation}
H_c=\sum_{k=1}^4\sum_{n \in {\Bbb Z}}
\Big[N^{k+}_n(\sqrt{(n+\delta)^2+f^2}+\beta \al p^+)
+N^{k-}_n(\sqrt{(n-\delta)^2+f^2}-\beta \al p^+)\Big]
+8\Delta(f;\delta),\label{closed H g}
\end{equation}
where $\Delta(f;\delta)$ is the zero point energy (see eq.(\ref{zero}))
and we defined the number operators as
\begin{equation}
\begin{split}
&N^{k+}_n = \left\{\begin{array}{ll}
\frac{1}{2w^+_n}\tilde 
\ap^{k}_{-n-\delta}\tilde {\bar \ap}^k_{n+\delta}
& \mbox{for} \quad n \geq -\delta \\
\frac{1}{2w^-_{-n}}\ap^{k}_{n+\delta}{\bar \ap}^k_{-n-\delta}
& \mbox{for} \quad n < -\delta
\end{array}\right.,\\
&N^{k-}_n = \left\{\begin{array}{ll}
\frac{1}{2w^-_n}\tilde {\bar \ap}^{k}_{-n+\delta}\tilde \ap^k_{n-\delta}
& \mbox{for} \quad n > \delta \\
\frac{1}{2w^+_{-n}}\bar \ap^{k}_{n-\delta}\ap^k_{-n+\delta}
& \mbox{for} \quad n \leq \delta
\end{array}\right..
\end{split}
\end{equation}
Notice that the zero point energy in (\ref{closed H g})
is canceled by the fermionic one due to the supersymmetry
\footnote{We can also obtain the total Hamiltonian by adding
the fermionic number operators $F_n^{k\pm}$ to $N_n^{k\pm}$
due to the Bose-Fermi degeneracy
(see \cite{HaTa} in detail).}.

Finally we would like to comment on the case of $p^+=0$. We need to
consider the case because in the closed string channel,
the Neumann boundary condition in the $x^+$ direction is
expressed as $\prt_\tau X^+=0$. In this case, we should modify
the $n=0$ sector in eq.(\ref{mode exp}) into the complex momenta
and positions, i.e., 
\begin{equation}
Z^k_0=(\al p^k_z \tau + z^k)e^{-i\delta \sigma}+\cdots,\quad
\bar Z^k_0=(\al \bar p^k_z \tau + \bar z^k)e^{i\delta \sigma}+\cdots,
\quad [z^k,\bar p^{k'}_z]=2i\delta_{kk'}.
\end{equation}

\subsubsection*{Open Strings on D1-brane}
Next let us consider open strings on the D1-brane
extended in the light cone directions
in the twisted PP-wave background. We have only to
consider the D1-brane at $Z^k=0$, because
the background is homogeneous one. We can obtain
the open string spectrum from the closed string one
using the doubling trick.
We need to set $\delta=0$ in eq.(\ref{mode exp})
because the light cone gauge
for the open strings on this D1-brane is $X^+=2\al p^+\tau$.
Noticing that the boundary condition $Z^k=0$ at $\sigma=0$ and $\pi$
is equivalent to the following gluing condition
\begin{equation}
\ap^k_n=-\tilde \ap^k_n \quad\mbox{for}
\quad n\neq 0,
\quad \ap^k_0=\tilde \ap^k_0=0,
\end{equation}
we obtain the mode expansion of the open string on the D1-brane as follows
\begin{equation}
Z^k_0=\sqrt{2\al}\sum_{n \neq 0} \frac{\ap^k_n}{w_n}\sin(n\sigma)
e^{-iw_n\tau},\quad w_n \equiv \mbox{sgn}(n)\sqrt{n^2+m^2},\quad
m \equiv 2\beta \al p^+.
\end{equation}
The commutation relations are
$[\ap^k_n,\ap^{k'}_\ell]=2w_n\delta_{\ell+n,0}\delta^{kk'}$.
The mode expansion of $\bar Z^k_0$ is just the complex
conjugate of $Z^k_0$.
The open string Hamiltonian is expressed as follows
\begin{equation}
{\cal H}_o=-2\al p^+p^-+H_{o}.
\end{equation}
The massive mode $H_o$ is given by
\begin{equation}
H_o=\sum_{n=1}^\infty \Big[N^{+}_n(\sqrt{n^2+m^2}
+m)+N^{-}_n(\sqrt{n^2+m^2}-m)\Big]+4\Delta(m;0)-2|m|,
\label{open H g}
\end{equation}
where we defined the number operators as follows
\begin{equation}
N^{+}_n =
\sum_{k=1}^4\frac{1}{2w_n}
\ap^{k}_{-n}{\bar \ap}^k_{n},\quad
N^{-}_n =
\sum_{k=1}^4
\frac{1}{2w_n}{\bar \ap}^{k}_{-n}\ap^k_{n}.
\end{equation}
The zero point energy $4\Delta(m;0)-2|m|$ is canceled by 
the fermionic one due to the supersymmetry as in the case of closed strings
\footnote{In contrast with the case of closed strings,
we also need the extra term
$h_o=-\sum_{k=1}^4\f{m}{2}(F^k_0-\f{1}{2})$ which comes from
the zero modes of the fermions on D1-brane (see \cite{GaGr} in detail)
as well as the fermionic number operators of massive modes
in order to obtain the total Hamiltonian.}.
Finally let us compute the following open string
vacuum amplitude
\begin{equation}
Z^k_{open}=\int_0^\infty \frac{dt}{t}
\int d p^+ d p^- \mbox{Tr}
\exp [-2\pi t (-2\al p^+p^- + H_{o})].
\end{equation}
When we define $q=e^{-2\pi t}$ as before, we find the non-trivial part
is given by
\begin{equation}
\mbox{Tr}e^{-2\pi t H_o}=\left[q^{\Delta(m;0)-\f{|m|}{2}}\prod_{p=1}^\infty
\frac{1}{(1-q^{\sqrt{p^2+m^2}+m})
(1-q^{\sqrt{p^2+m^2}-m})}\right]^4
\equiv f(q,m)^4.
\end{equation}

\subsubsection*{Open Closed Duality of D1-brane}
In the rest of this section, we would like to study
the open closed duality of D1-brane in the twisted PP-wave
background. Noticing that the relation between the mass parameter
of the open strings $m$ and that of the closed strings $\delta$
is $mt=-i\delta$ (see eq.(\ref{example})), we obtain
the following modular transformation rule
\footnote{In the limit $\delta \rightarrow 0$, this modular property
reduces to that of the $\eta$ function, i.e.,
$\eta(it)^{2}=t^{-1}\eta(is/\pi)^{2}$.}
\begin{equation}
\begin{split}
f(q,m)=
\sqrt{\frac{2\pi^3\ap^{'2} \sin (2\pi \delta)}{\delta}}
\f{\tilde q^{\Delta(i\delta;\delta)}}{(2\pi \al s)}\prod_{n\neq 0}\frac{1}
{1-\tilde q^{\sqrt{(n+\delta)^2-\delta^2}}}
\equiv \tilde f(\tilde q,\delta), \quad \tilde q= e^{-2s},
\end{split}\label{D0 cardy}
\end{equation}
where we used the modular properties of the massive theta functions
\footnote{We used the following modular transformation
of the massive theta function (see appendix \ref{sec:massive})
\begin{equation}
\Theta_{(0,-\delta)}(t;m)=\Theta_{(\delta,0)}(1/t,mt).
\end{equation}
After setting $m^2=
\frac{-\delta^2+ \epsilon^2}{t^2}$
and taking the limit $\epsilon \rightarrow 0$, we
obtain eq.(\ref{D0 cardy}).}
in the PP-wave background. Therefore the Cardy's condition
is given by (see eq.(\ref{cardy}))
\begin{equation}
\langle D1,w|e^{-sH_c}|D1,w\rangle=
\tilde f(\tilde q,\delta_w)^4,\quad \delta_w \equiv \beta Rw,
\end{equation}
where $|D1,w\rangle$ denotes the boundary state for D1-brane
with the winding $w$ sector.
We can construct the boundary state for the D1-brane satisfying
this condition. The result is
\begin{equation}
|D1,w\rangle = \left(\frac{2\pi^3\ap^{'2}
\sin (2\pi \delta_w)}{\delta_w}\right)
\int \f{d p_z^k d \bar p_z^k}{(2\pi)^8}
|w \rishi
\otimes|p_z^k,\bar p^k_z\rangle,
\end{equation}
where $|w\rishi$ is the following Ishibashi state for the D1-brane
\begin{equation}
|w \rishi \equiv \exp\Big[\sum_{\substack{n>\delta_w \\ n\neq 0}}\f{1}{2w_n^+}
\bar {\tilde \ap}^{k}_{-n+\delta_w}\ap^k_{-n+\delta_w}
+\sum_{\substack{n>-\delta_w \\ n\neq 0}}\f{1}{2w_n^-}
{\tilde \ap}^{k}_{-n-\delta_w}\bar \ap^k_{-n-\delta_w}
\Big]|0\rangle. \label{D1 ishi}
\end{equation}

\subsection{Boundary State for the Supertube on G\"odel Universe}
In the previous section we have reviewed the string theory
on the twisted PP-wave background which is the T-dual of
the type IIA G\"odel universe. We have also constructed
the boundary state for the D1-brane at $Z^k=0$ in the twisted
PP-wave background.

We now would like to construct the boundary state for a moving
spiral D1-brane in the twisted PP-wave background using the null
deformations as in the case of the flat spacetime.
This spiral D1-brane corresponds to a supertube in the G\"odel universe.
In this section we will consider only the case of $k=1$ for simplicity.

As in the case of the flat spacetime, 
let us consider the following boundary state without
light gauge fixing
\begin{equation}
|N,1\rangle=e^{-\f{i}{2\pi \al}\int_0^{2\pi}d\sigma [\rho\cos(\f{Nx^+}{R})P_1
+\rho\sin(\f{Nx^+}{R})P_2]}|D1\rangle,\label{godel tube bs}
\end{equation}
where $P_\mu \equiv G_{\mu\lambda}\prt_\tau X^\lambda$ is the canonical
momentum of the world sheet theory
on the twisted PP-wave background (\ref{twist ppw}). 
We can also replace $P_i$ with $\prt_\tau X^i$ in eq.(\ref{godel tube bs})
noticing the Neumann condition $\prt_\tau X^+|D1\rangle=0$.
Formally we can show that
the above boundary state satisfies boundary conditions for
the $(N,1)$ D-helix (\ref{tube bc})
\footnote{Null deformations keep (GS) fermionic boundary conditions unchanged
(see \cite{SkTa1} in detail). Therefore fermionic deformation
operators are not needed in eq.(\ref{godel tube bs}).}
and the boundary conformal invariance
by using the canonical quantization
$[P_\mu(\sigma,\tau),X^\nu(\sigma',\tau)]=
-i\delta_\mu^\nu\delta(\sigma-\sigma')$. 
However we should take a great care since the naive expression
(\ref{godel tube bs}) would be divergent. We can avoid
the divergence by renormalization, however it breaks 
the boundary conformal invariance in general.

Fortunately in the case of the twisted PP-wave background,
the canonical quantization condition indicates that the operator
products $X^+X^+$ and $X^+X^i$ have no singular terms.
As in \cite{CaKl}, this implies that the deformation operator
is exactly marginal \footnote{It is hard to show
its exact marginality because the string theory on the background
can not be solved exactly without light cone gauge fixing.}. 
Therefore the boundary state (\ref{godel tube bs})
is that for the $(N,1)$ D-helix up to a normalization factor.
The normalization factor is determined by the Cardy's condition and
we will find later that no extra normalization factor is needed as in
the case of the flat spacetime.
The same argument shows that we can construct the boundary states for null
deformed D-branes on the twisted PP-wave background
by multiplying those for static D1-branes by
the Wilson line like operators \footnote{In the GS formalism,
the vertex operator for the gauge field $A^i$ is given by
\begin{equation}
V_B(k)=(\prt_\tau X^i-\frac{1}{4}S^a\gamma^{ij}_{ab}S^bk^j)
e^{ik \cdot X}.
\end{equation}
Therefore there is no fermionic contribution for $k^+=k^i=0$.
The same argument can be done for transverse scalars
(the T-dual of gauge fields).
}. Notice that deformations with respect to $X^-$ are not allowed as expected
because the operator product $X^-X^i$ will have singular terms.

Now in the light cone gauge $X^+=Rw\sigma$,
eq.(\ref{godel tube bs}) is expressed as the following form
\begin{equation}
|N,1 \rangle = e^{-\f{i}{2\pi\al}\int_0^{2\pi}d \sigma V_{N,1}}|D1\rangle,
\quad V_{N,1}=\f{\rho}{2}
[e^{-iNw\sigma}\prt_\tau Z^1 +e^{iNw\sigma}\prt_\tau \bar Z^1].
\end{equation}
We can compute the vacuum amplitude between the same D-helix
in the same way as in the
case of the flat spacetime. The result is given by
\begin{equation}
{\cal A}(s)=\sum_w
\exp \Big[-\frac{\rho^2}{\al}w_{-Nw}\tanh\left(\f{sw_{-Nw}}{2}\right)\Big]
e^{-\frac{R^2w^2s}{2\al}}\tilde f(\tilde q,\delta_w)^4,
\label{vac supertube}
\end{equation}
where we defined 
\begin{equation}
w^2_{-Nw}=(\beta wR)^2\Big[(-\frac{N}{\beta R}+1)^2-1\Big]
=(wR)^2\Big(\f{1-2b\beta}{b^2}\Big).
\end{equation}
In the final line we used the relation $b=\f{R}{N}$. Notice that
$w_{-Nw}$ becomes pure imaginary for $2b\beta>1$
After taking the T-duality in the $y$ direction
$(R \rightarrow \al/\tilde R)$, we obtain the vacuum amplitude
of the $(N,1)$ supertube.

Let us consider the naive 
$s \rightarrow 0$ limit, the UV limit of closed strings.
In this limit, the non-trivial factor in eq.(\ref{vac supertube}) 
can be interpreted as the following radius shift
\begin{equation}
\exp\Big[-\frac{R^2w^2s}{2\al}-\frac{\rho^2w^2_{-Nw}s}{2\al}\Big]
=\exp\Big[-\frac{R^{'2}w^2s}{2\al}\Big],
\end{equation}
where the shifted radius $R^\prime$ is given by
\begin{equation}
R^{'2}=R^2\Big[
1+\f{\rho^2}{b^2}(1-2b\beta)
\Big].
\end{equation}
For $2b\beta >1$, the square of shifted radius becomes negative
when $\rho^2 > \f{b^2}{2b\beta -1}$. Therefore one might think
this is the tachyonic instability of closed strings wound
in y direction (or closed strings with the nonzero momentum $p_y$
in the G\"odel universe). However we should interpret
the region $s\sim 0$ as the open string IR rather than the closed string UV.
Notice that the naive $s\rightarrow 0$ limit is different from
the IR limit of open strings because the conformal invariance of the
world sheet theory is broken by the gauge fixing, i.e.,
the mass parameter is changed under the open closed duality.
\subsection{Open String Spectrum on Supertube in G\"odel Universe}
As we have found in the previous section, we have to treat the vacuum
amplitude in the open string picture to carry out the correct physical
interpretations. In this section we would like to rewrite the obtained
vacuum amplitude in the language of the open string channel using
the same method as before.
As in the case of the flat spacetime, we will compare
the result to the following open string metric of the
supertube in G\"odel universe
\begin{equation}
G^{tt}_{eff}=-\left(1+\rho^2\f{(1-2b\beta)}{b^2}\right).\label{eff godel}
\end{equation}
Especially, we are interested in the case of $2b\beta >1$
because the effective metric $G^{tt}_{eff}$ changes
its sign for large values of $\rho$ in the case 
\footnote{Although one might think the $G^{tt}_{eff}$
becomes the same as the original metric $G^{tt}$ for $b\beta=1$,
we can not consider the case because the radius
$\rho$ goes to infinity in the case.}.
This is the signal of
the closed timelike curves in the G\"odel universe. Therefore
we will consider only the case of $2b\beta >1$ below.

Using the correspondence $2i\al p^+ t\leftrightarrow wR$,
we can rewrite the non trivial part in eq.(\ref{vac supertube})
in the language of the open string channel as follows
\begin{equation}
-\frac{\rho^2}{\al}w_{Nw}\tanh \frac{w_{Nw}s}{2}
\rightarrow 
-2\pi t \Big[\frac{\rho^2 p^+}{\pi}\sqrt{\f{2b\beta-1}{b^2}}
\tanh\Big(\pi \al p^+\sqrt{\f{2b\beta-1}{b^2}}\Big)\Big].
\label{oc dual godel}
\end{equation}
Then we obtain the following open string Hamiltonian of the $(N,1)$ D-helix 
\begin{equation}
{\cal H}_o=-2\al p^+p^- +\frac{\rho^2 p^+}{\pi}\sqrt{\f{2b\beta-1}{b^2}}
\tanh\Big(\pi \al p^+\sqrt{\f{2b\beta-1}{b^2}}\Big)
+H_{o}.
\end{equation}
Notice that the open-closed duality (\ref{oc dual godel}) shows
that no extra normalization factor in the boundary state
(\ref{godel tube bs}) is needed as we promised.
After taking T-duality in y direction, we obtain the open string Hamiltonian
on the $(N,1)$ supertube on the G\"odel universe. For open strings
with no windings in the $\tilde y$ direction, the result is given by
\begin{equation}
{\cal H}_o=-\al E^2 +\frac{\rho^2 E}{\pi}\sqrt{\f{2b\beta-1}{b^2}}
\tanh\Big(\pi \al E\sqrt{\f{2b\beta-1}{b^2}}\Big)
+H_{o},\label{godel open sp}
\end{equation}
where the massive mode $H_o$ is given by
\begin{equation}
H_o=\sum_{n=1}^\infty \Big[N^{+}_n(\sqrt{n^2+(2\beta \al E)^2}
+2\beta \al E)+N^{-}_n(\sqrt{n^2+(2\beta \al E)^2}-2\beta \al E)\Big].
\end{equation}
Remember that we can neglect the zero point energy due to the
supersymmetry.
In the low energy limit $|\al E \sqrt{\f{2b\beta-1}{b^2}}| \ll 1$, 
we can read off the open string metric (\ref{eff godel}) from
the Hamiltonian (\ref{godel open sp}) as expected.

Now let us consider the open string spectrum ${\cal H}_0=0$.
If there were no $\al$ correction in eq.(\ref{godel open sp}),
this would have the form $-\al G^{tt}_{eff}E^2=H_o$.
Therefore we can interpret the effect
of the closed timelike curve as the tachyonic instability of
the open strings. That is, the energy $E$ becomes imaginary. 
However there is a non trivial $\al$ correction in eq.(\ref{godel open sp}).
Since the magnitude of $\tanh x$ is always smaller than that of $x$,
this tachyonic instability becomes smaller at the string theory level
than obtained from the low energy interpretation. Notice that we can
indeed find the real $E$ satisfying ${\cal H}_o=0$ for any radius of
the supertube $\rho$
\footnote{Noticing the following behaviour of the function ${\cal H}_0$
at $E=0$ and $|E|\rightarrow \infty$
\begin{equation}
{\cal H}_o(E=0)=\sum_{n=1}^\infty n(N^+_n+N^-_n)>0,\quad
{\cal H}_o(|E|\rightarrow \infty)\rightarrow -\infty,
\end{equation}
we find one solution of ${\cal H}_o=0$ in the range of $E>0$ and
another one in $E<0$.
}. Therefore we can conclude that there is no tachyonic instability
of open strings at the string theory level.
\section{Conclusion}
\setcounter{equation}{0}
\hspace{5mm}
In this article we have constructed the boundary states for supertubes in
the flat spacetime in the covariant formalism and read stringy information
from the vacuum amplitude of open strings on the supertube. The supertube
is the stable tubular D2-branes with the `critical' electric
flux and constant magnetic flux on them, i.e., the
bound state of $N$ D2-branes, $k$ D0-branes and $q$ F-strings. Supertubes
are also the BPS solitons preserving 8 supercharges. Taking
the T-duality in the direction in which the supertube is extended,
we obtain the moving spiral D1-brane (D-helix) wound $k$ times in the
T-dual direction and $N$ times in the angular direction. We can also
interpret this D-helix as the symmetrically placed $k$ segments of
spiral D1-branes. Since adding the
`critical' electric flux corresponds to null boosting in the T-dual picture,
we can obtain the D-helix from a static D1-brane via null deformation,
the deformation only in the $x^+$ direction. This is
consistent with the fact that any null deformed D-brane preserves 8
supercharges which
is the same as the number of the supercharges of the supertube.
This is the basic idea
needed to construct the boundary states.

First we have constructed the boundary state
for the $(N,k)$ D-helix 
from $k$ segments of D1-branes via null deformation
in the covariant formalism. We have found that we can
interpret one segment of the spiral D1-brane as a `fractional'
brane and the $(N,k)$ D-helix as a `bulk' brane in an orbifold theory.
We have also computed the vacuum amplitude of the open strings
on the supertube in an exact form in the closed string channel.
In the superstring case, we found the vacuum amplitude is zero
as expected from the point of view of the supersymmetry.

In particular, we found the open string spectrum on the $k=1$
supertube in an exact form using the open closed
duality in the discretized light cone gauge.
This is due to the fact that the $k=1$ D-helix has the good periodicity
in the light cone direction. In the low energy limit,
we can read the open string metric from the open string spectrum.
On the other hand we have found a non-trivial $\al$ correction
in the high energy region. We found there is a `critical' energy
for open strings on the supertube
due to this stringy correction.

It is known that we can deform the section of the
supertube preserving 8 supercharges. We can easily
understand this fact in the T-dual picture from the
point of view of null deformation.
We have also constructed the boundary states for these
deformed supertubes in the same way. We can exactly
calculate the vacuum amplitude which vanishes
in the superstring case and find an exact open string spectrum
in the case of $k=1$.

We have also considered supertubes in the supersymmetric
type IIA G\"odel universe in order to use them as probes
of closed timelike curves. Since the T-dual of this universe
is the type IIB PP-wave background with maximal supersymmetry
where null deformation is allowed, we can construct the boundary
states for supertubes in the same way as in the case of the flat
spacetime. We calculated the vacuum amplitude for the
$k=1$ supertube and found the open string spectrum. 
In the open string spectrum, the effect of the closed timelike
curve appears as the tachyonic instability of open strings.
We found this tachyonic instability disappears
at the string theory level due to the non-trivial $\al$ correction
of the spectrum.
\subsection*{Acknowledgments}
\hspace{5mm}
The author would like to thank T. Eguchi, K. Hashimoto, Y. Hyakutake,
K. Sakai, Y. Sugawara, T. Takayanagi
and S. Yamaguchi for useful discussions and comments. The author
is also grateful to Yukawa Institute for Theoretical Physics for
hospitality during the completion of this paper.
\vskip2mm

\appendix
\section{Massive Theta Functions}
\label{sec:massive}
\setcounter{equation}{0}
\hspace{5mm}
Here we would like to summarize the definition
of the massive theta functions \cite{Ta} and their modular
properties.
We define the massive theta functions as follows 
\begin{equation}
\Theta_{(a,b)}(\tau;m)\equiv e^{4\pi \tau \Delta(m;a)}
\prod_{n \in {\Bbb Z}}(1-e^{-2\pi \tau \sqrt{m^2+(n+a)^2}+2\pi ib})
(1-e^{-2\pi \tau \sqrt{m^2+(n+a)^2}-2\pi ib}),
\end{equation}
where the zero-point energy $\Delta(m;a)$ is given by
\begin{equation}
\begin{split}
\Delta(m;a) &\equiv \frac{1}{2}\Big(
\sum_{n \in {\Bbb Z}}\sqrt{m^2+(n+a)^2}-\int_{-\infty}^\infty
dk \sqrt{m^2+k^2}
\Big)\\
&=-\frac{1}{2\pi^2}\sum_{n=1}^\infty \int_0^\infty ds e^{-sn^2-
\frac{\pi^2 m^2}{s}}\cos (2\pi na).\label{zero}
\end{split}
\end{equation}
These theta functions have the following modular property
(see \cite{Su} in detail)
\begin{equation}
\Theta_{(a,b)}(1/t,m t)=\Theta_{(b,-a)}(t;m).
\end{equation}

\end{document}